\documentclass[fleqn,twoside]{article}
\usepackage{espcrc2}
\usepackage{epsbox}
\usepackage{epsf}
\renewcommand{\epsilon}{\varepsilon}

\newcommand{\f}{\frac}

\newcommand{\tri}{\triangle}
\newcommand{\m}{\mbox}

\title{Numerical study 
on the correlation between CP violation\\
in neutrino oscillations and baryogenesis}
\author{T. Endoh, T.
Morozumi and~A.
Purwanto\thanks{ On leave
from Jurusan Fisika FMIPA-ITS Surabaya, and Universitas Muhammadiyah Malang
Indonesia.}\\ 
Department of Physics, Hiroshima University, Higashi Hiroshima - 739-8526, Japan}
\begin{document}

\begin{abstract}
We numerically study the correlation between CP violation in the
neutrino oscillations and baryogenesis in the seesaw model. In this
 study we get the heavy Majorana neutrino masses and lepton number
 asymmetries from their decays by fitting the data of
 neutrino oscillations and by working on some hypothesis of the
 Dirac-Yukawa term for neutrinos.
\end{abstract}

\maketitle
\section{Introduction}
In the previous paper[1] we have identified the CP violating
phases of the seesaw
model and found six independent phases.
In the case where the
heavy right-handed Majorana mass has a strong hierarchy in the diagonal basis
$M_1~{\gg}~M_2~{\gg}~M_3$, the diagonalization of the seesaw matrix
is possible under the assumption that the elements in Yukawa-Dirac
mass matrix are rather universal.
Among six phases, three of them are identified as
a Dirac phase and two Majorana phases in the light neutrino sector while
the remaining three arise from the mixing of the light neutrinos and
heavy neutrinos.
Other result is that three phases of the low energy
sector in principle can be determined by using neutrino oscillation
and
double beta decay
experiments. The remaining three appears in the lepton number
asymmetry of heavy Majorana neutrinos decays. Under the assumption,
it was shown
the phases in low energy sector
and the phases in high energy sector are not related each others and the
correlation between CP violation at low energy and CP violation
at high energy is rather weak.

In this paper, we explore more general case.
Following the arguments given by [2], we assume that 
the Yukawa-Dirac mass term is rather hierarchical and their magnitudes
are given by up type quark masses. By reducing numbers of independent
parameters, we can study the case that the  correlation does exist. The purpose of this paper is to study numerically the correlation
between  CP violating phases in
MNS matrix and CP violating phases for baryogenesis through leptogenesis.

\section{Seesaw Model and MNS Matrix}
 The mass Lagrangian of the seesaw model has the form
\begin{equation}
{\cal L}_{m}= -\f{1}{2}({\nu}_{L},  N_{R}^c)M_{\nu}\left(\begin{array}{ccc}
 {\nu}_{L}^c \\
 N_{R}
\end{array}
\right) + h.c., \label{2}
\end{equation}
where $M_{\nu}$ is $6{\times}6$ mass matrix
\begin{equation}
 M_{\nu}=\left(\begin{array}{ccc}
0 & m_{D} \\
m_{D}^T & M
\end{array}
\right), \label{3}
\end{equation}
where $m_D$, $M$ are Dirac and Majorana mass matrix respectively.
This $M_{\nu}$ can be diagonalized by unitary matrix $V$,
i.e., $m_{d}=V^TM_{\nu}V$, where $V$ is given as \ $
  \left(
 {\nu}_{L} \ \
 N_{R}^c
\right)^T = V^{*}{\nu}_{L}^{\alpha}$.

In the non-decoupling case [1]and by using the mass eigenstate for charged lepton, the $V_{MNS}$ matrix is the submatrix
$3{\times}6$ of $V$ itself. 
General analysis on this extended $3{\times}6$ MNS matrix results in six independent CP phases.

Rewriting $m_d$ in the form $\m{diag}(m_{\nu},\  M_d)$, 
$V$ in the submatrices K, R, S, T, and the order of S and R as given in[3],
then the diagonalization we will obtain a usual relation of the seesaw model
\begin{equation}
 m_{\nu}=-K^{\dagger}m_{D}\f{1}{M}m_{D}^{T}K^{*}.
\end{equation}

Further analysis gives 
the charged current which shows  
that $K$ and $R$ give the charged current couplings of charged
leptons to the light neutrinos ${\nu}_L$ and to the heavy neutrinos
$N_R$, respectively. In the decoupling limit, $R$ can be
neglected and only $3{\times}3$ MNS matrix $K$ is relevant. 
\section{Correlation between the Phases}
We consider the diagonalization of Eq.(3), and for calculational purpose
we take 
all six phases of $3{\times}3$ MNS
matrix appear. We parameterize
\begin{equation}
m_{D}=UY_{\tri},
\end{equation} 
where $U$ is the unitary matrix and $Y_{\tri}$ is triangle matrix whose
real diagonal elements $y_{i}$ and complex off-diagonal elements $y_{ij}$
where $i>j$. To make a connection between MNS matrix and lepton number
asymmetry we set $U=1$ and suppress six independent parameters. With
this working hypothesis, from the low energy neutrino oscillation data,
we can determine the scales of heavy Majorana neutrinos and the lepton
number asymmetries from their decays.

Without loose of generality we take matrix $M$ is real diagonal,
$M=\m{diag}(M_{1},\ M_{2},\ M_{3})$. Expansion Eq.(3) in component $(K
m_{\nu}K^{T})_{ij}= (-Y_{\tri}\f{1}{M}Y_{\tri}^{T})_{ij}$, we have
\begin{eqnarray}
M_{1}&=& \f{1}{m_{11}}y_{1}^{2}, \nonumber  \\
M_{2}&=& \f{m_{11}}{m_{11}m_{22}-m_{12}^{2}}y_{2}^{2}, \\ \label{16}
M_{3}&=& \f{m_{11}m_{22}-m_{12}^{2}}{det|K m_{\nu}K^{T}|}y_{3}^{2}, \nonumber
\end{eqnarray}
and
\begin{eqnarray}
y_{21}&=&\f{m_{12}}{m_{11}}y_{1}, \nonumber \\
y_{31}&=&\f{m_{13}}{m_{11}}y_{1}, \\ \label{17}
y_{32}&=&\f{m_{11}m_{23}-m_{12}m_{13}}{m_{11}m_{22}-m_{12}^{2}}y_{2},
\nonumber
\end{eqnarray}
where $m_{ij}=(K m_{\nu}K^{T})_{ij}$.
The Eq.(5) leads to three constraints for six phases of general unitary MNS
matrix, hence the MNS matrix has only three free phases as we
demand. Whereas the Eq.(6) gives relation
between the phases in low and high energy
sector. 

Because both $M_{i}$ and $y_{i}$ are real then we can easily see that
Eq.(5) gives simple constraints i.e. each denominator of these
equations must be real. By  
using the standard parameterization of the CKM matrix [4],
we parameterize the MNS matrix $K$ of the
form $K = \m{diag}
\left( e^{i{\beta}_{1}},  \   e^{i{\beta}_{2}}, \   e^{i{\beta}_{3}}
\right)V_{CKM}({\delta}) \m{diag}\left(
1,  \  e^{i{\alpha}_{1}},  \ e^{i{\alpha}_{2}}
\right) $.
These $K$, $m_{\nu}$ and 
the above 
constraints yield
${\beta}$ as function of neutrino
masses $m_i$, mixing angles ${\theta}_{ij}$ and CP phases ${\delta},
{\alpha}_{1},{\alpha}_{2}$.
These
${\beta}_{i}$
are not independent phases and can be absorbed by charged lepton fields,
and only Dirac phase ${\delta}$ and majorana phases ${\alpha}_{1}
, \ {\alpha}_{2}$  are independent
phases. Next, the reality condition of $M$ will give
$M_{i}=M_{i}(y_{i},m_{1},m_{2},m_{3},{\theta}_{12},{\theta}_{13},
{\theta}_{23,},{\delta},{\alpha}_{1},{\alpha}_{2})$
and therefore, it is possible to estimate their values using the neutrino
data on
neutrino masses, mixing angles and some assumption on
the lightest
$m_{1}$ and diagonal elements $y_{i}$. For numerical calculation, we
will set $y_1= O(m_u),\ y_2= O(m_c)$ and $y_3= O(m_t)$.

To determine the correlation between the phases in high energy
 sector live in $y_{ij}$ and the phases in low energy sector i.e,
 ${\delta}$, ${\alpha}_{1}$, and ${\alpha}_{2}$ we factor out
 the phases from $y_{ij}$ as $|y_{ij}|e^{i{f}_{ij}}$.
This factorization and taking into account the reality of 
 $y_i$ as well as the previous constraints also give the simple
 constraints i.e. multiplication $exp({-i{f}_{ij}})$ to each numerator
 of $y_{ij}$ in Eq.(6) must be real.
\section{Numerical Results}
Take the values of
neutrino masses $m_{3}= 5.5{\times} 10^{-2} ~\m{eV}, \ \ m_{2}=
4.5{\times} 10^{-3} ~\m{eV}$, and $
m_{1}= 10^{-2}{\times} m_{2}$ and mid values of mixing angles [5] with all phases are zero
\begin{eqnarray}
K=\left(\begin{array}{ccc}
0.80643 & 0.585905 & 0.0799147  \\
-0.474852 & 0.561102  &  0.677997 \\
0.352402 & -0.584705 &  0.730707
\end{array}
\right).
\end{eqnarray} \label{32}
\begin{figure}[h]
\begin{minipage}{0.7\hsize}
\begin{center}
\epsfile{file=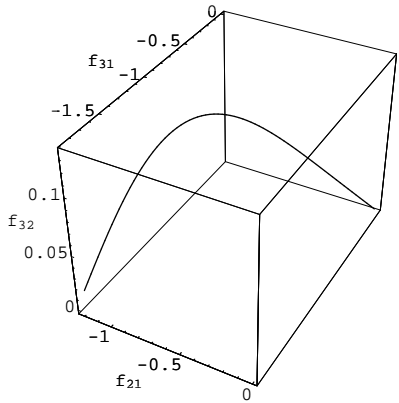,width=\hsize}
\caption{\small The correlation between $f_{ij}$ and ${\delta}$. We set
 ${\alpha}_{i}=0$.}
\end{center}
\end{minipage}
\end{figure}
\begin{figure}[htbp]
\begin{tabular}{c}
\begin{minipage}{0.8\hsize}
\begin{center}
\epsfile{file=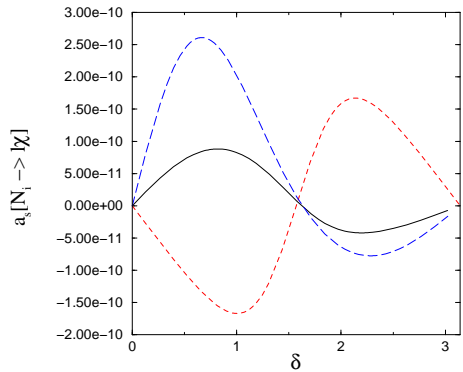,width=\hsize}
\end{center}
\end{minipage}
\end{tabular}
\caption{The asymmetries $a_{s}$ as function of ${\delta}$. Dashed,
 long-dashed and solid line are $a_{s1}, a_{s2}/10$ and $a_{s3}/10^4$
 respectively. We set ${\alpha}_{i}=0$.}
\end{figure}
These values are combined with $ y_1 = 0.003 ~\m{GeV},  \ \ y_2 = 1.25
~\m{GeV}, \ \ y_3 = 170 ~\m{GeV}.$
After substituting them into Eq.(5), we get the masses of
right-handed neutrino,  
\begin{eqnarray}
M_{1} &=& 4.67{\times} 10^{6}   ~~~\m{GeV}, \nonumber \\
M_{2} &=& 9.49{\times} 10^{10} ~~~\m{GeV}, \\ \label{34}
M_{3} &=& 8.22 {\times}10^{16} ~~~\m{GeV}. \nonumber
\end{eqnarray}
An example of correlation between CP phases in low energy sector (${\delta}$)
and CP phases for leptogenesis
($f_{21}$, $f_{31}$, $f_{32}$) is shown in
Figure 1. We change ${\delta}$ from $0$ to ${\pi}$.

Using the above results we evaluate the lepton number asymmetry coming from
the decay of massive right-handed Majorana neutrino $N_i$ into charged
leptons $l$ and Higgs fields ${\chi}$
\begin{eqnarray}
a_s(N_i {\rightarrow}l^{\mp}{\chi}^{\pm})&=& \f{v^{-2}}{4{\pi}(y_{\tri}^{\dagger}y_{\tri})^{ii}}{\sum}_{k}\m{Im}[(y_{\tri}^{\dagger}y_{\tri})
^{ik}]^2 \nonumber \\
&{\times}&\left[\m{I}\left( \f{{M_k}^2}{{M_i}^2}\right)+\f{{M_i}^2}{{M_i}^2-{M_k}^2}\right].
\end{eqnarray}
where $v=246 ~\m{GeV}$.
The dependence of asymmetry on low CP phases can be obtained, and the
dependence on Dirac phase ${\delta}$ is given by Figure 2. From this
figure,  we can see the order of asymmetry from $N_{R1}$ is $10^{-10}$, and
the asymmetry of the heavier ones are larger.

\section{Conclusions}
In the present paper, we study the correlation between CP
violation at low energy and at high energy.
Under the working hypothesis  on the Dirac mass term of
neutrino sector, we can determine CP violating phases
which are related to baryogenesis through leptogenesis from the three CP
violating phases in MNS matrix, the neutrinos mixing angles and their mass
differences. The correlation between the lepton number asymmetry
and the CP violating phases in MNS matrix is numerically studied.
We estimate the masses and the decay widths
of three heavy Majorana neutrinos. These are sufficient for prediction
of the baryon number asymmetry. 

\vspace{2mm}
{\bf Acknowledgment}\par 
We would like to thank organizers of KEKTC5. This work is
 supported by the Grand-in-Aid for scientific research No.13640290 from
 the Ministry of Education; Science and Culture of Japan.

\end{document}